\documentclass[letterpaper,twocolumn,10pt]{ilass}
\def\thepage{}

\usepackage{amssymb}
\usepackage{amsmath}
\usepackage{subfig}
\usepackage[x11names]{xcolor}
\usepackage{tikz,tikz-3dplot,tkz-euclide}
\usetikzlibrary{calc}
\usetikzlibrary{arrows}
\usetikzlibrary{shapes.geometric, arrows, intersections, through}
\usetikzlibrary{decorations.text, decorations.shapes, backgrounds}
\usetikzlibrary{arrows.meta}
\usetikzlibrary{colorbrewer}
\usetikzlibrary{patterns}
\usepackage{pgfplots} 
\usepackage{pgfplotstable}
\pgfplotsset{compat=1.10} 
\pgfplotsset{width=6cm} 
\pgfplotsset{height=3.6cm} 
\pgfplotsset{grid style={solid}}
\tikzset{math3d/.style={z={(-0.353cm,-0.353cm)},y={(0cm,1cm)},x={(1cm,0cm)}}}
\usepgfplotslibrary{colorbrewer}
\usepgfplotslibrary{colormaps}
\usepgfplotslibrary{patchplots}
\usetikzlibrary{patterns}
\usetikzlibrary{pgfplots.groupplots}

\usepgfplotslibrary{external} 
\usepackage{adjustbox}
\usepackage{ifpdf}
\usepackage[round]{natbib}

\title{\vspace{-0.25in} {\small \em ~} \newline \newline
        \large {\bf Simulating interfacial flows: a farewell to planes} }
\author{\large F. Evrard\footnote{Corresponding Author: fabien.evrard@cornell.edu}~$^{\dagger\star}$, R. Chiodi$^{\ddagger}$, B. van Wachem$^{\star}$, and O. Desjardins$^{\dagger}$ \\~\\
\large $^{\dagger}$Cornell University, Ithaca, NY 14853, USA\\
\large $^{\star}$Otto-von-Guericke-Universit\"at, Magdeburg, D-39106, Germany\\
\large $^{\ddagger}$Los Alamos National Laboratory, Los Alamos, NM 87545, USA\\
        \large }

\date{ \normalsize  \centerline{\bf Abstract} \vspace{0.05in}
 \begin{minipage}{6.5in}
 \normalsize Over the past decades, the volume-of-fluid (VOF) method has been the method of choice for simulating atomization processes, owing to its unique ability to discretely conserve mass. Current state-of-the-art VOF methods, however, rely on the piecewise-linear interface calculation (PLIC) to represent the interface used when calculating advection fluxes. This renders the estimated curvature of the transported interface zeroth-order accurate at best, adversely impacting the simulation of surface-tension-driven flows.
 In the past few years, there have been several attempts at using piecewise-parabolic interface approximations instead of piecewise-linear ones for computing advection fluxes, albeit all limited to two-dimensional cases or not inherently mass conservative. In this contribution, we present our most recent work on three-dimensional piecewise-parabolic interface reconstruction and apply it in the context of the VOF method. As a result of increasing the order of the interface representation, the reconstruction of the interface and the estimation of its curvature now become a single step instead of two separate ones. The performance of this new approach is assessed both in terms of accuracy and stability and compared to the classical PLIC-VOF approach on a range of canonical test-cases and cases of surface-tension-driven instabilities.
 \end{minipage} \vspace{-0.25in}
}

\begin{document}

\ifpdf
\DeclareGraphicsExtensions{.pdf, .jpg}
\else
\DeclareGraphicsExtensions{.eps, .jpg}
\fi

\maketitle

\clearpage 
\pagenumbering{arabic}
\setcounter{page}{2}

\section*{Introduction} 
Simulating liquid atomization is a notoriously difficult task, since it requires numerical methods that both provide strict mass conservation and a robust estimation of the interface curvature, so as to accurately predict the evolution of the gas-liquid interface as well as the size and velocity distributions of the dispersed droplets. Among the numerical methods available for modeling fluid-fluid interfaces, Moment-of-Fluid~(MOF) \citep{Dyadechko2005} and Volume-of-Fluid~(VOF) \citep[e.g.,][]{Owkes2014} methods are among the only ones able to guarantee strict mass conservation, making them the methods of choice for simulating atomization processes. MOF and VOF methods typically rely on piecewise-linear interface calculations (PLIC) for solving the transport of the liquid phase's indicator function, which has the direct consequence of limiting the accuracy of the estimated interface curvature to zeroth-order at best. As a result, the modeling of surface tension bears a significant error that cannot be mitigated by sheer computing power (i.e., by refining the computational mesh). 

A natural solution for escaping this deadlock is to increase the order of representation of the piecewise interface approximations used for transporting the phase indicator function, for instance using piecewise-parabolic interface calculations (PPIC) instead of piecewise-linear ones (PLIC). The algorithmic complexity associated with the cutting of computational cells by paraboloids instead of planes, however, has long been a barrier for accomplishing this. The tools previously at the disposal of the community for calculating the moments of a polyhedron clipped by a paraboloid, for instance based on numerical quadrature \citep[e.g.,][]{Chierici2022} or order reduction \citep{Renardy2002}, have either been too expensive or too inaccurate. 

In this work, we present a three-dimensional PPIC-based VOF method that is both strictly mass conservative and computationally efficient. More specifically, we introduce the building blocks necessary for such implementation, that are: 1) The solution of the forward problem, i.e., calculating the fluid moments in a computational cell or any arbitrary polyhedron from the knowledge of a paraboloid interface approximation; 2) The solution of the backward problem, i.e., reconstructing an optimal piecewise-parabolic interface approximation from the knowledge of fluid moments; 3) The advection schemes that can be used for transporting the phase indicator function; 4) The procedure employed for transforming the surface tension force distribution into a volumetric momentum source. We study the order of accuracy of the proposed numerical framework, both for the transport of the phase indicator function and the estimation of curvature. Finally, we assess the accuracy and robustness with which surface tension is modeled, in the context of the multiphase flow solver \texttt{NGA2}\footnote{https://github.com/desjardi/NGA2}.

\section*{Forward problem}
\begin{figure}[h]
        \begin{center}
        \subfloat[Dodecahedron clipped by a paraboloid and the projection of its edges on the $xy$-plane.]{\parbox{3.0in}{\centering\vspace{-2mm}\includegraphics[width=2.0in]{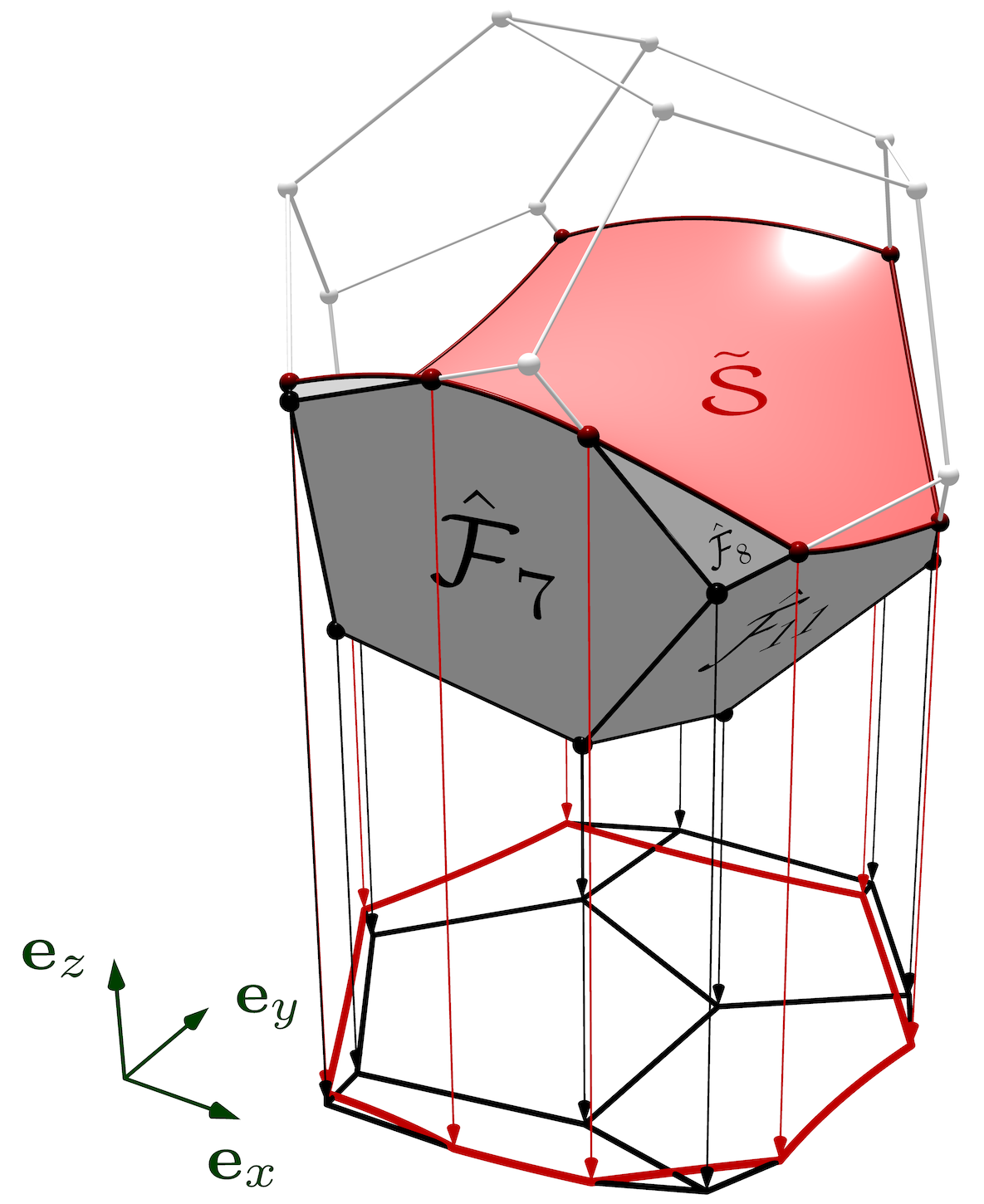}}}\\
        \subfloat[Edges of the projected clipped faces, on which the 1D moment integrals are calculated. They either consist of line segments or conic section arcs.]{\includegraphics[width=3.0in]{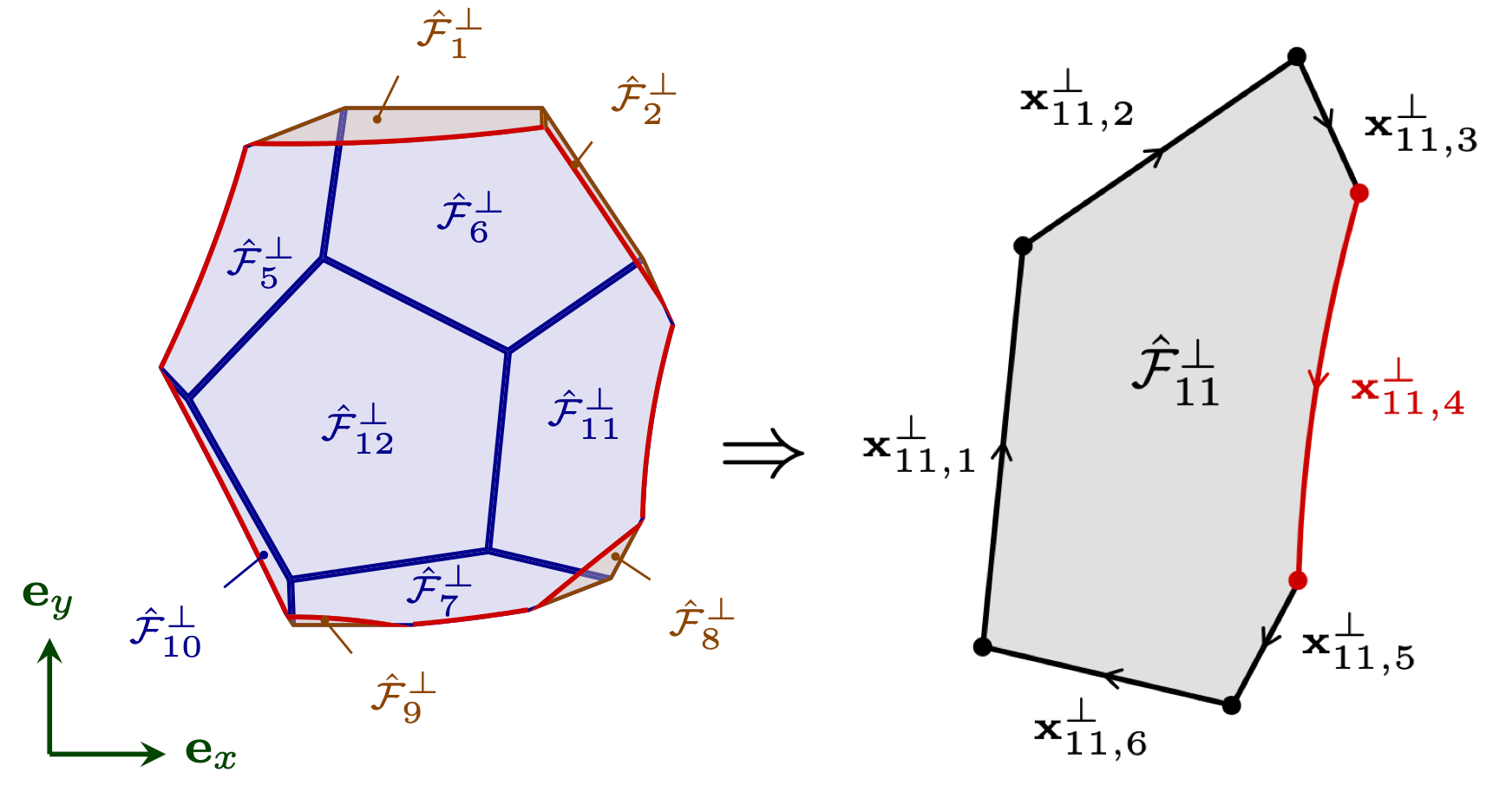}}
        \end{center}
        \caption{Illustration of the domains on which the moments are integrated for the forward problem \citep{Evrard2022}.}
        \label{fig1} 
\end{figure}

\begin{figure}[h] \centering
        \includegraphics{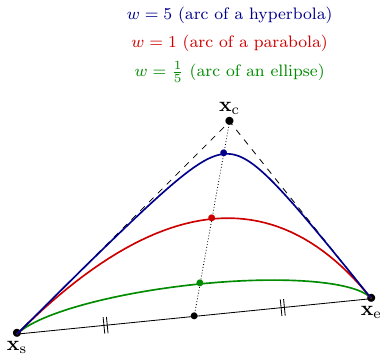}
        \caption{Three rational B\'ezier curves linking a start point $\smash{\mathbf{x}_{\text{s}}}$ to an end point $\smash{\mathbf{x}_{\text{e}}}$. Also shown are the control point $\smash{\mathbf{x}_{\text{c}}}$ and weight $\smash{w}$.}
        \label{fig:bezierarc}
\end{figure}

\begin{figure} \centering
        \adjincludegraphics[width=0.18\linewidth,trim=0 0 0 0,clip=true]{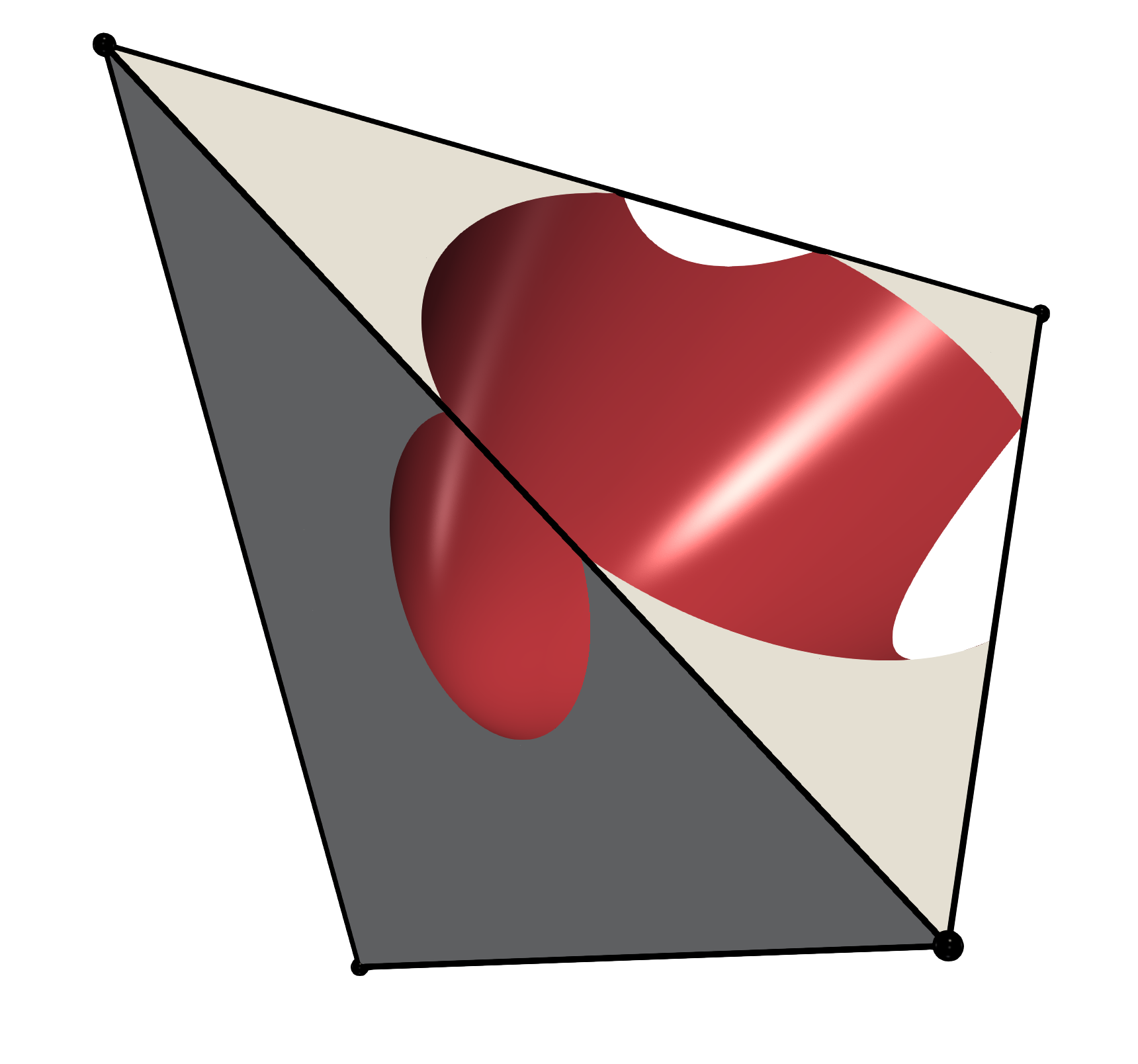}
        \adjincludegraphics[width=0.18\linewidth,trim=0 0 0 0,clip=true]{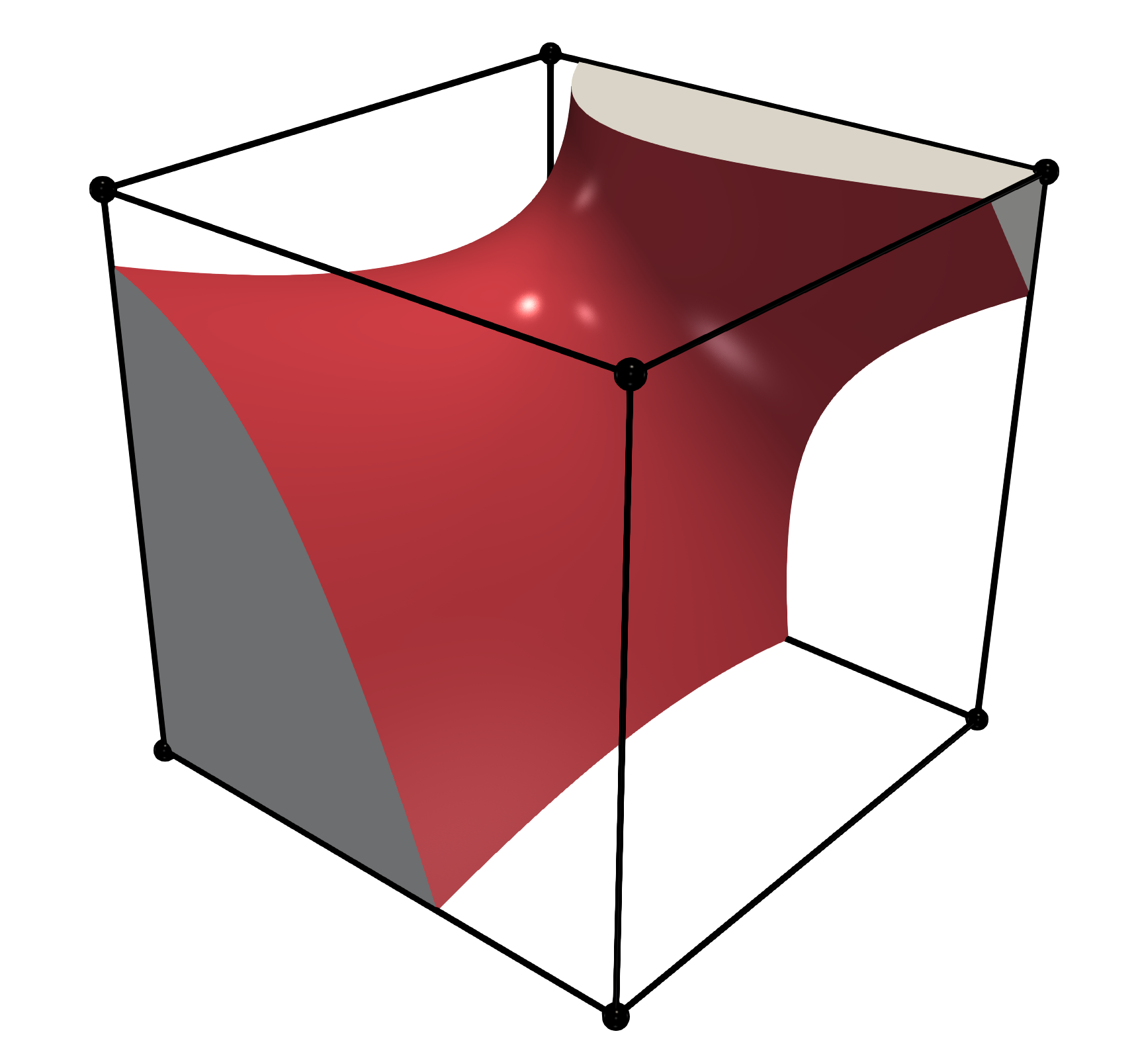}
        \adjincludegraphics[width=0.18\linewidth,trim=0 0 0 0,clip=true]{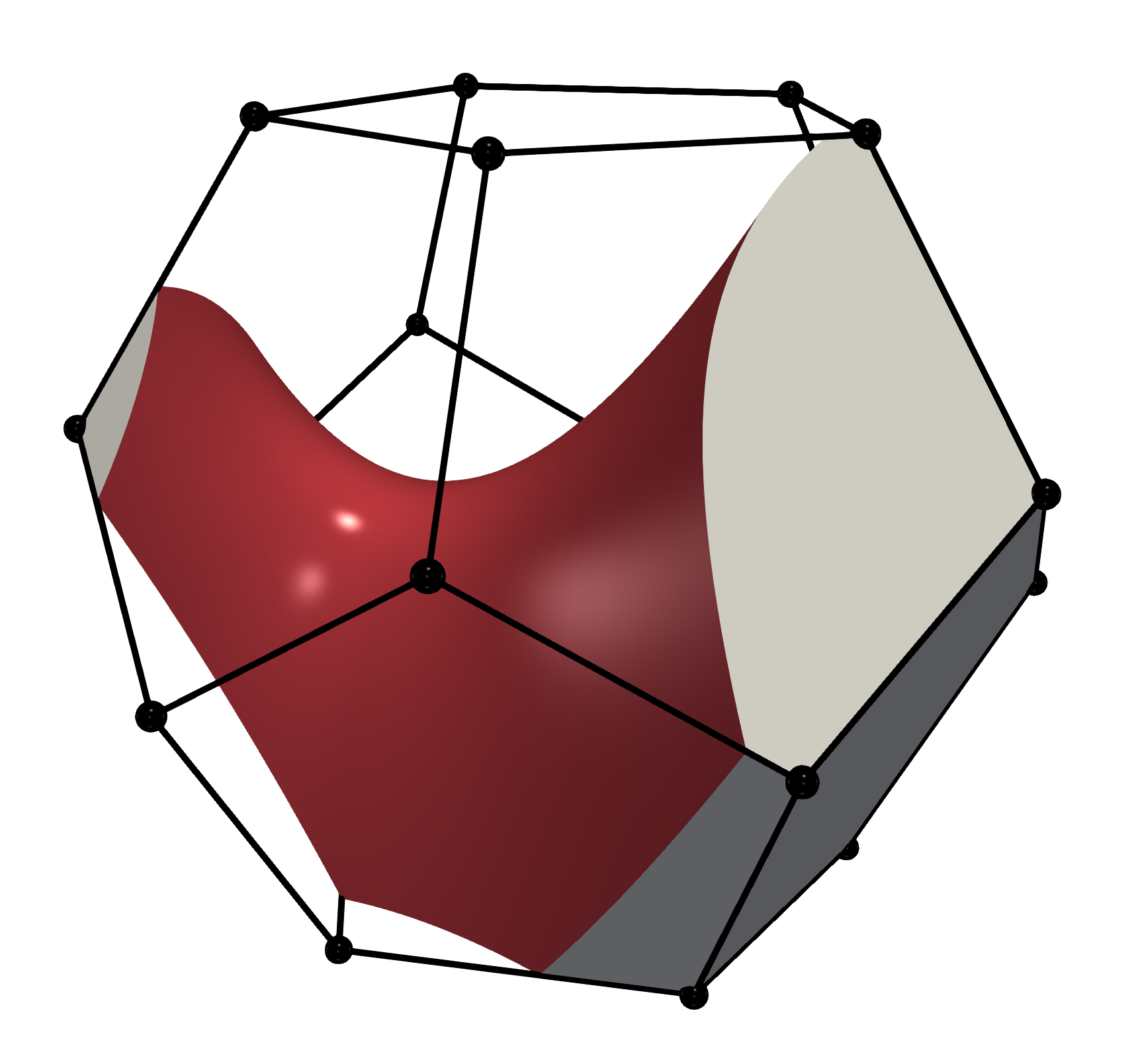}
        \adjincludegraphics[width=0.18\linewidth,trim=0 0 0 0,clip=true]{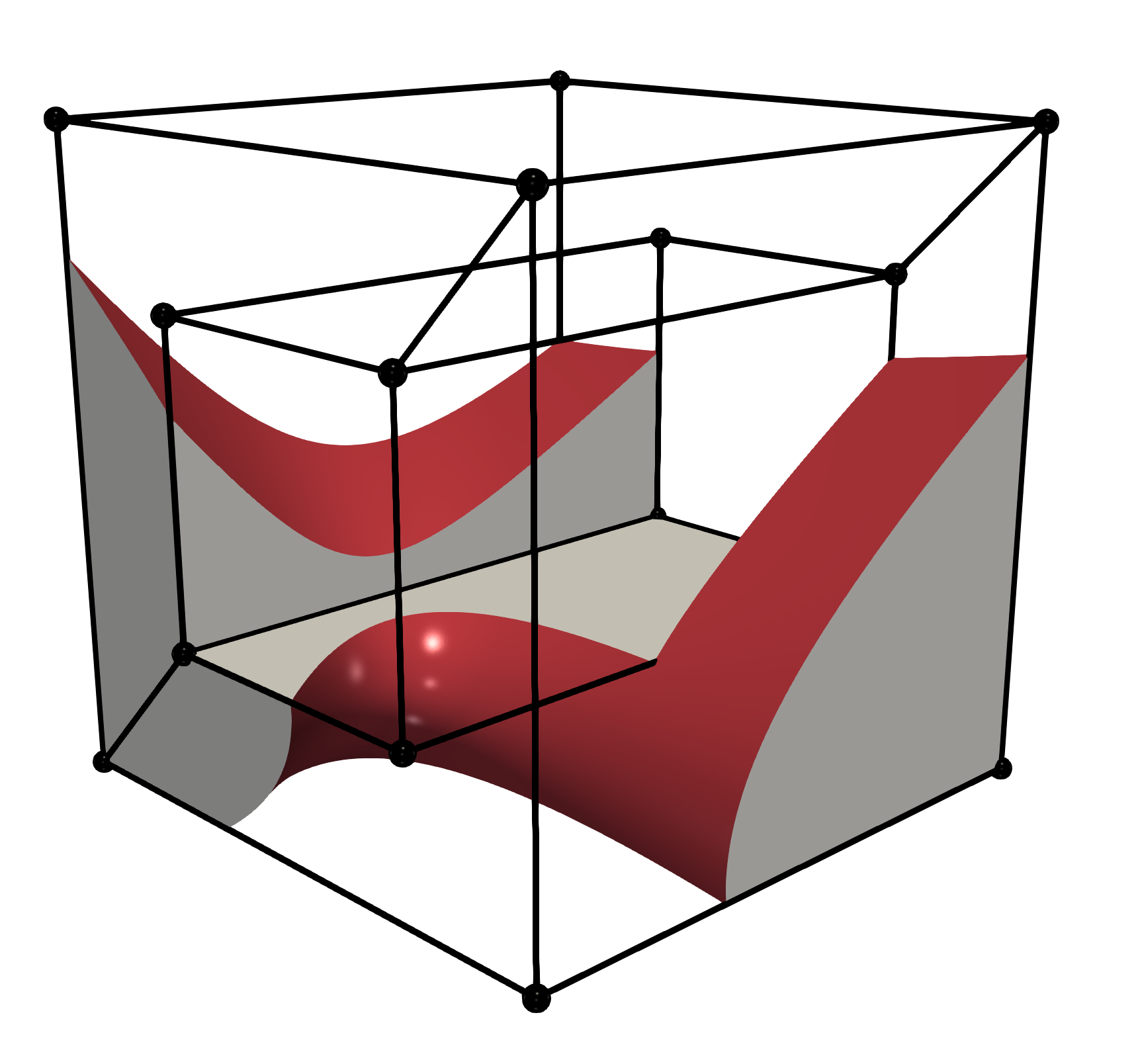}
        \adjincludegraphics[width=0.18\linewidth,trim=0 0 0 0,clip=true]{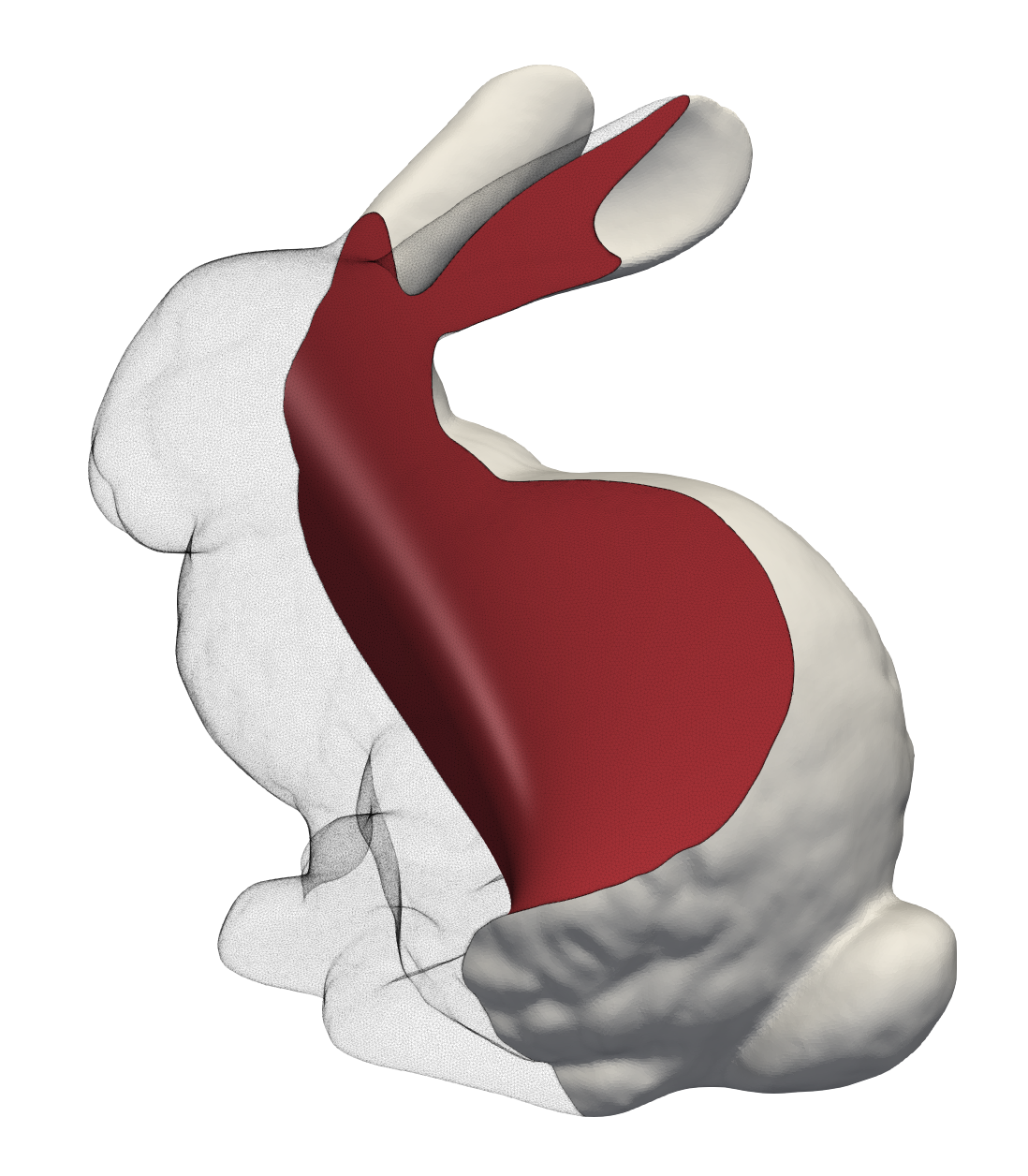}
        \caption{Shapes on which our forward problem solution algorithm has been tested \citep{Evrard2022}.}
        \label{fig3}
\end{figure}

The solution of the forward problem, i.e., the calculation of the moments of any polyhedron clipped by a paraboloid, uses the closed-form expressions of \citet{Evrard2022}. These expressions have been obtained from successive applications of the divergence theorem, transforming the 3D integrals of the moment monomials over the clipped polyhedron into a sum of 1D integrals over the oriented projected edges of the clipped polyhedron, as illustrated in Figure~\ref{fig1}. The integration domains of these 1D integrals consist of line segments and conic section arcs. 
The parametrization of the latter is in the form of rational B\'ezier curves given as
\begin{equation}
	\mathbf{x}(t) = \dfrac{B_0(t) \, {\mathbf{x}}_{\text{s}} + w \, B_1(t) \, {\mathbf{x}}_{\text{c}} + B_2(t) \, {\mathbf{x}}_{\text{e}}}{B_0(t) + w \, B_1(t) + B_2(t)} \, , \label{eq:rat_bezier}
\end{equation}
with the Bernstein polynomials
\begin{align}
	B_0(t) & = (1-t)^2 \, ,\\
	B_1(t) & = 2(1-t)t \, ,\\
	B_2(t) & = t^2 \, ,
\end{align}
i.e., each conic section arc is parametrically defined as a rational polynomial for the parameter $t \in [0,1]$ with coefficients that are functions of the start-point $\mathbf{x}_{\text{s}}$, control-point $\mathbf{x}_{\text{c}}$, and end-point $\mathbf{x}_{\text{e}}$ of the arc, as well as of the rational B\'ezier weight $w$ related to the nature of the conic section. Examples of such rational B\'ezier curves are shown in Figure~\ref{fig:bezierarc}.
This provides two main advantages compared to a parametrization using trigonometric functions: 1)~The same parametrization can be used seamlessly for the case of an elliptic, parabolic, or hyperbolic conic section; 2)~Substantial round-off errors due to the use of floating-point arithmetics are prevented. 
These closed-form expressions have been tested over a range of shapes (shown in Figure~\ref{fig3}) and for more than $200$ million intersection configurations. Overall, the moment estimations have been shown to be machine-accurate, robust, and only about 6 times more expensive than for single plane cutting \citep{Evrard2022}. The code for solving this forward problem is openly available in the \texttt{Interface Reconstruction Library}\footnote{https://github.com/robert-chiodi/interface-reconstruction-library/tree/paraboloid\_cutting}.

\section*{Backward problem}
\begin{figure} \centering\vspace{-2mm}
        \subfloat[PLIC]{\includegraphics[width=0.49\linewidth]{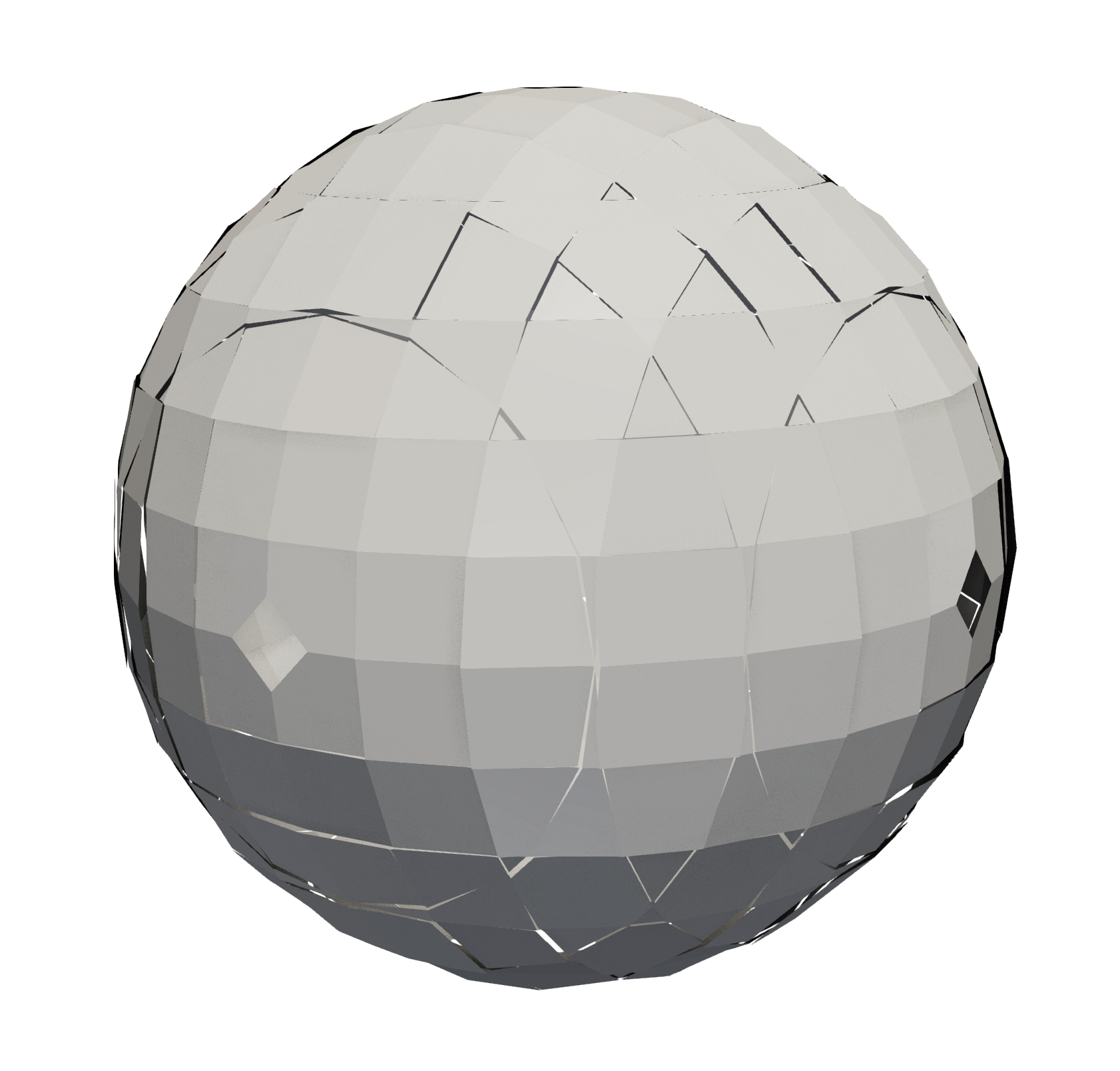}}
        \subfloat[PPIC]{\includegraphics[width=0.49\linewidth]{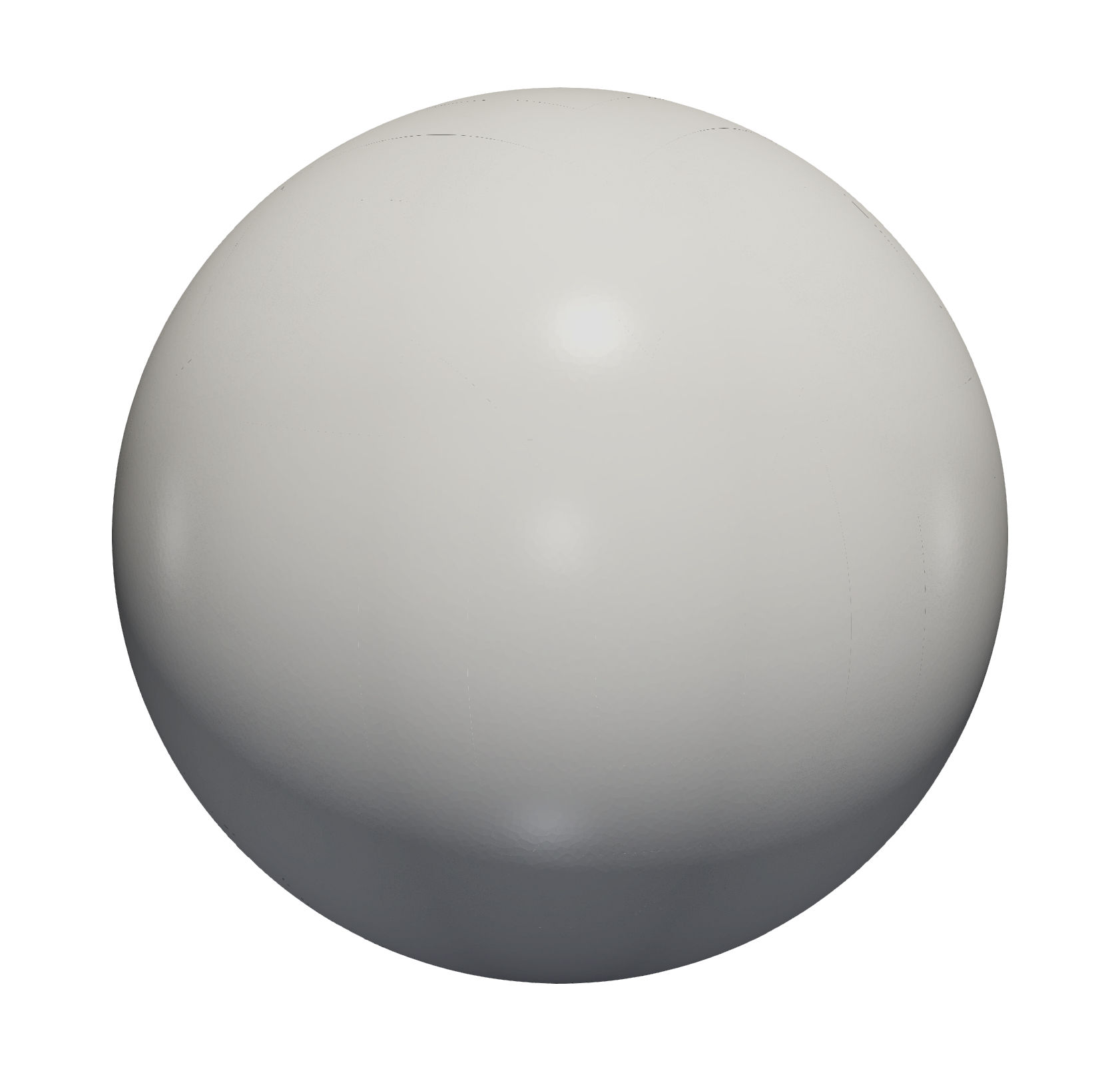}}\vspace{-2mm}
        \caption{PLIC and PPIC of a sphere with $10$ cells/diameter using the methods of \citet{Pilliod2004} and \citet{Jibben2019}, respectively.}
        \label{fig4}
\end{figure}

The solution of the backward problem, i.e., the reconstruction of a paraboloid approximation of the interface using the fluid moments, is based on the paraboloid fitting algorithm of \citet{Jibben2019}. This procedure consists in minimizing the volume between the PLIC in a local neighborhood of computational cells and the local paraboloid approximation. Defining a local frame of reference, whose $z$-basis-vector points towards the approximate local normal direction of the interface, it can be formulated as the minimization problem
\begin{equation}
        \min\limits_\mathcal{S} \;\sum_{n} \left(\int_{\mathcal{P}^{\perp}_n} \!\!\!\!\!\left(f_{\mathcal{P}_n}(x,y) - f_\mathcal{S}(x,y)\right) \ \mathrm{d}x\,\mathrm{d}y \right)^2  
\end{equation}
where $\sum_{n}$ is to the sum over all computational cells of the local neighborhood, $\mathcal{P}^{\perp}_n$ is the projection of the PLIC of the $n^\text{th}$ cell in that neighborhood on the $xy$-plane of the local frame of reference, $f_{\mathcal{P}_n}(x,y)$ is the parametrization of the PLIC of the $n^\text{th}$ cell in the local frame of reference, and $f_\mathcal{S}(x,y)$ is the parametrization of the seeked parabolic approximation in the local frame of reference. This problem takes the form of a linear least-square problem whose solution is that of a linear system of equations \citep{Jibben2019}. In the present work, the PLIC is first obtained using the LVIRA method of \citet{Pilliod2004} before conducting the parabolic fit. In a final step, the fitted paraboloid $\mathcal{S}$ is translated along the local $z$-direction so as to match the local volume fraction with machine-accuracy, using the bisection method.
An example of the PLIC and resulting PPIC obtained for a spherical interface is shown in Figure~\ref{fig4}.

\section*{Advection}
Once solution procedures for the forward and backward problems described in the previous sections are available, the actual transport of the phase indicator function can employ traditional advection schemes as previously developed for PLIC-based flux calculations. In the present work, we consider the operator-split advection scheme of \citet{Weymouth2010} as well as the semi-Lagrangian unsplit advection scheme of \citet{Owkes2014} using both PLIC and PPIC. Both schemes transport the phase indicator function while retaining strict conservation and boundedness of the volume fractions.

\section*{Surface tension}
To model surface tension, we employ the Continuum-Surface-Force (CSF) model of \citet{Brackbill1992}. The main difference between a PLIC- and a PPIC-based approach here lies in the fact that the PPIC contains local curvature information, while the PLIC does not. Using PPIC, it is then possible to directly compute the local interface curvature from the solution of the backward problem, so a separate curvature estimation step is not required. In the current work, curvature is integrated over the parabolic surface inside each computational cell, using the same parametrization as for the forward problem, and based on the surface integration procedure detailed in \citet{Evrard2022}. 

\section*{Numerical tests}
\begin{figure}\centering
              \includegraphics{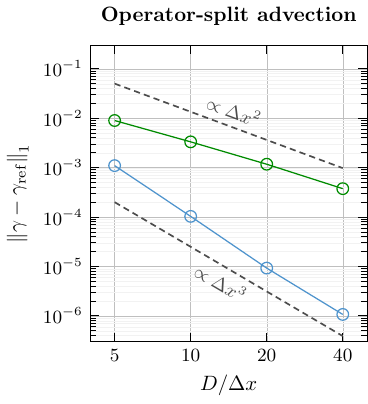}\quad\includegraphics{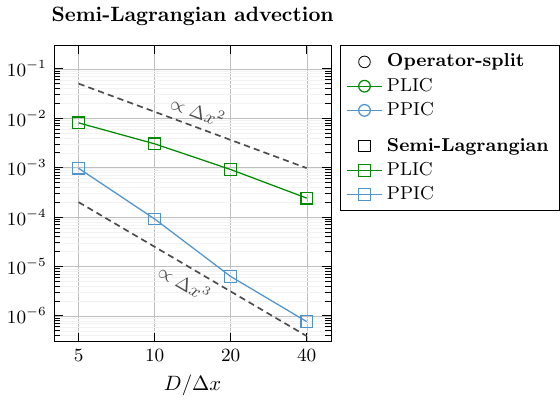}
              \caption{Shape errors for the case of the uniform translation of a sphere with diameter $D$, using PLIC and PPIC as well as operator-split and semi-Lagrangian advection schemes.}
              \label{fig5}
\end{figure}

\begin{figure}\centering
              \includegraphics{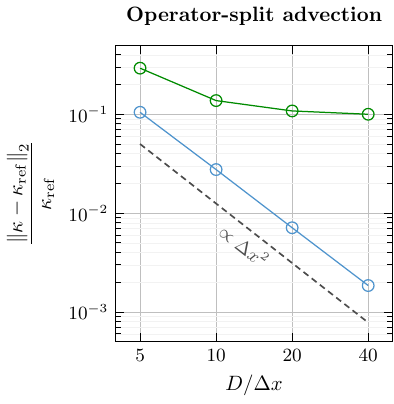}\quad\includegraphics{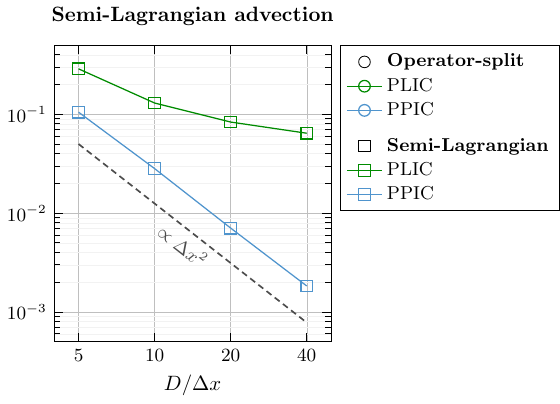}
              \caption{Curvature errors for the case of the uniform translation of a sphere with diameter $D$, using PLIC and PPIC as well as operator-split and semi-Lagrangian advection schemes.}
              \label{fig6}
\end{figure}

The proposed numerical framework for interfacial flow modeling with the PPIC-VOF method is implemented in the open-source multiphase flow solver \texttt{NGA2}. In a first instance, the accuracy of the phase indicator transport is studied for a case of uniform translation of a sphere in a periodic box. The direction of the uniform velocity is chosen to be not aligned with the dominant grid directions, and the timestep is chosen based on a $\text{CFL} = 0.5$. The sphere is transported across a distance corresponding to approximately $7.5$ sphere diameters. The convergence of the shape error is shown in Figure~\ref{fig5}, demonstrating the expected $2^\text{nd}$ and $3^\text{rd}$ order convergence rates of the PLIC and PPIC approaches, respectively. Figure~\ref{fig6} shows the convergence of the estimated curvature of the interface, post transport. Using PLIC, the expected $0^\text{th}$ convergence rate is recovered. Using PPIC, on the other hand, the curvature errors converge with mesh refinement, with a $2^\text{nd}$ order rate. Note that, in the case of PLIC, the curvature of the interface is estimated using the procedure of \citet{Jibben2019} after the interface has been transported.

\begin{figure} \centering
      \includegraphics{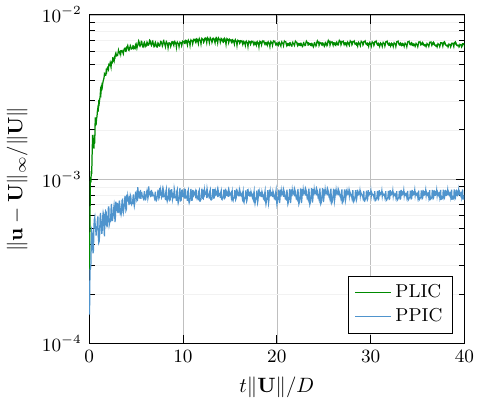}
      \caption{Parasitic flow currents for the case spherical interface with diameter $D$ subject to uniform translation with the velocity $\mathbf{U}$. The Laplace number for this case is $\text{La = 12000}$.}\label{fig7}
\end{figure}

In a second instance, we study the influence of the choice of PLIC or PPIC on the production of parasitic/spurious flow currents, a well known issue in interfacial flow modeling \citep{Abadie2015}. The setup is identical to the previous test case, with the difference that surface tension is now turned on. The Laplace number for this test case is $\text{La = 12000}$, the sphere is resolved by $12.8$ cells across its diameter, the density and viscosity ratios are chosen equal to~$1$, and the timestep is chosen so as to satisfy the capillary timestep constraint \citep{Denner2015}. The maximum magnitude of the observed parasitic flow currents, normalized by the magnitude of the uniform translational velocity, is plotted in Figure~\ref{fig7}. These results show that, all things being equal apart from the type of surface approximation, the PPIC-based VOF approach produces parasitic flow currents that are about $10$ times less pronounced than the PLIC-based approach. 

\section*{Summary}
In this contribution, we propose a Volume-of-Fluid (VOF) method that relies on piecewise-parabolic interface calculations (PPIC) instead of piecewise-linear interface calculations~(PLIC). The forward problem of estimating the first moments of any polyhedron clipped by a paraboloid is solved using the closed-form expressions of \citet{Evrard2022}. The backward problem of reconstructing a paraboloid from the knowledge of fluid moments follows the fitting procedure of \citet{Jibben2019}. The transport of the phase indicator function uses classical split or unsplit geometrical advection schemes, and the interface curvature is extracted directly from the solution of the backward problem through its integration over the clipped paraboloid surface. The resulting framework produces convergent curvature estimates for transported interfaces, a feature that PLIC-based approaches cannot provide, by design. Moreover, the use of PPIC for simulating interfacial flows generates parasitic/spurious flow currents that are an order of magnitude less pronounced than with the classical PLIC-VOF method.

\section*{Acknowledgements}
This project has received funding from the European Union's Horizon 2020 research and innovation programme under the Marie Skłodowska-Curie grant agreement No 101026017 \raisebox{-.125\height}{\includegraphics[height=2ex]{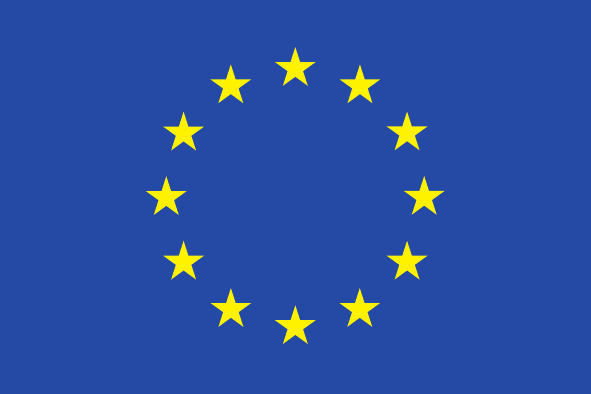}}\,.

\onecolumn

\twocolumn

\bibliographystyle{ilass}

 \end{document}